\documentclass[aps,prc,twocolumn,superscriptaddress,amsfonts,amsmath,amssymb,showpacs,floatfix,nofootinbib]{revtex4-1}

\usepackage{graphicx}
\usepackage{longtable}
\usepackage{bm}
\usepackage{multirow}
\usepackage{hyperref}
\usepackage{ulem}
\hypersetup{colorlinks,citecolor=blue,filecolor=blue,linkcolor=blue,urlcolor=blue}
\usepackage{color}
\usepackage{dcolumn}

\begin{document}

\title{Effect of Finite Binding on the Apparent Spin-Orbit Splitting in Nuclei}

\author{B.~P.~Kay}
\email{kay@anl.gov}
\affiliation{Physics Division, Argonne National Laboratory, Argonne, Illinois 60439, USA}
\author{C.~R.~Hoffman}
\affiliation{Physics Division, Argonne National Laboratory, Argonne, Illinois 60439, USA}
\author{A.~O.~Macchiavelli}
\affiliation{Nuclear Science Division, Lawrence Berkeley National Laboratory, Berkeley, California 94720, USA}

\date{\today}

\begin{abstract}

The apparent splitting between orbitals that are spin-orbit partners can be substantially influenced by the effects of finite binding. In particular, such effects can account for the observed decrease in separation of the neutron $1p_{3/2}$ and $1p_{1/2}$ orbitals between the $^{41}$Ca and $^{35}$Si isotopes. This behavior has been the subject of recent experimental and theoretical works and cited as evidence for a proton ``bubble'' in $^{34}$Si causing an explicit weakening of the spin-orbit interaction. The results reported here suggest that the change in the separation between the $1p_{3/2}$ and $1p_{1/2}$ partners occurs dominantly because of the behavior of the energies of these $1p$ neutron states near zero binding.

\end{abstract}

\pacs{}

\maketitle

To describe the ordering of levels in nuclei and the pattern of magic numbers a spin-orbit interaction had to be added to the nuclear Hamiltonian which is the basis of the very successful nuclear shell-model~\cite{Mayer}. The magnitude of the spin-orbit interaction is determined empirically and is not fully understood quantitatively. The spin-orbit coupling must be a surface term, and is usually included in the one-body mean field as a potential proportional to the derivative of the nuclear density.

Recent works have postulated the presence of a proton ``bubble'' in $^{34}$Si~\cite{Burgunder14,Mutschler17}, suggesting that the central depletion in the proton density results in an `interior' contribution to the spin-orbit interaction, opposite in sign to that of the outer surface, causing a factor of approximately two reduction in the spin-orbit splitting of the neutron $1p$ orbitals. In this work we show that the low binding energy of the neutron states considered, quantitatively accounts for the reduction. The effects of finite binding must be taken into account before other explanations are considered.

\begin{figure}
\centering
\includegraphics[scale=0.80]{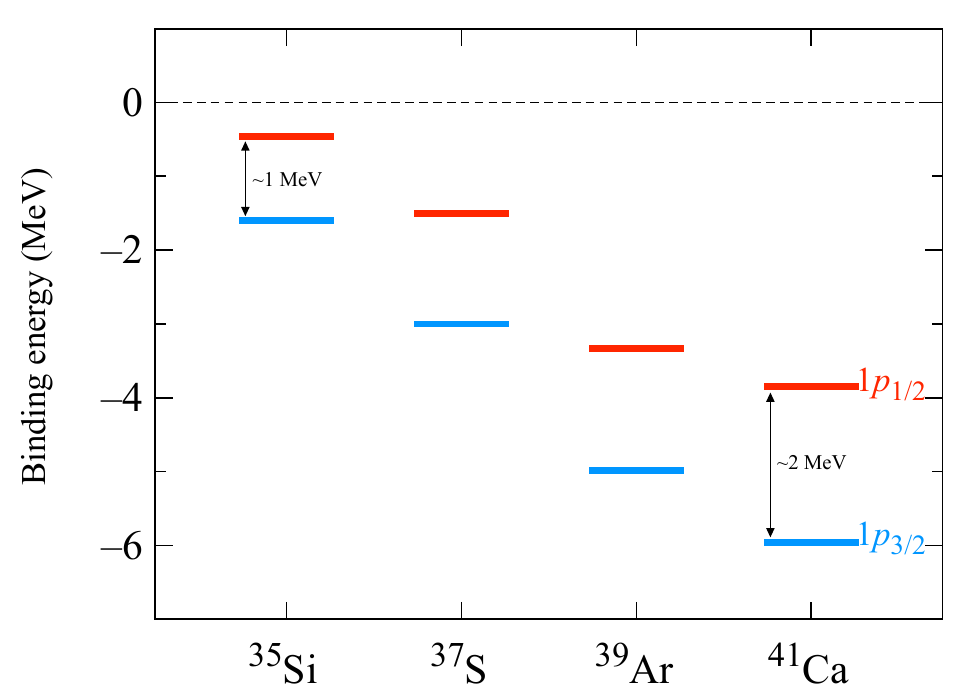}
\caption{\label{fig1} Experimentally determined binding energies~\cite{nndc,note} of the $1p_{3/2}$ and $1p_{1/2}$ single-particle (centroid) energies for $^{35}$Si, $^{37}$S, $^{39}$Ar, and $^{41}$Ca.}
\end{figure}

Data~\cite{nndc} from neutron-adding ($d$,$p$) reactions on $^{40}$Ca, $^{38}$Ar, $^{36}$S, and most recently $^{34}$Si using a radioactive ion beam~\cite{Burgunder14}, provide information on the location of the $1p_{3/2}$ and $1p_{1/2}$ single-particle strength outside of the $N=20$ closed neutron shell. This information is reasonably complete. Using the dominant fragments of the neutron $1p_{3/2}$ and $1p_{1/2}$ strength as a measure of the single-particle energies it was shown~\cite{Burgunder14} that the spin-orbit splitting between the neutron $1p_{3/2}$ and $1p_{1/2}$ states decreases by almost 1 MeV from $^{41}$Ca to $^{35}$Si, with most of that decrease happening between $^{37}$S and $^{35}$Si. To first order the use of the dominant fragments is a fair approximation, although the actual centroids, which can be determined from the available experimental data~\cite{note} (and see Supplemental Information~\cite{sm}), lie at slightly different energies. In Fig.~\ref{fig1}, we show the binding energy of the $1p_{3/2}$ and $1p_{1/2}$ centroids. Immediately it can be seen that the $1p$ orbitals move closer to each other, with the separation changing smoothly by about 1~MeV as $Z$ decreases. There is no experimental evidence for an abrupt change between S and Si for the centroids. The $1p_{1/2}$ state is close to the separation threshold at $^{35}$Si.

%---------------------------------------------------FIGURE 1------------------------------------------------------
%-----------------------------------------------------------------------------------------------------------------------

It was recently emphasized~\cite{Hoffman14,Hoffman16} that bound states with low angular momentum, especially $s$ states, linger below threshold, showing a reluctance to become unbound. Such effects had already been noted by Bohr and Mottelson~\cite{Bohr69}. The bulk features of the dramatic variations in the separation of the proton and neutron $1s$ and $0d$ orbitals between He and O can be attributed to weak-binding effects on the $s$ states, with the tensor and spin-orbit components of the residual two-body proton-neutron interaction accounting for only a small fraction of the total change~\cite{Hoffman14,Hoffman16}. 

The `lingering' effect means that the rate at which the eigenstate moves with changing radius of the potential, changes as the state approaches zero binding. Hamamoto and Sagawa~\cite{Hamamoto04} explored the splitting between the \mbox{$1p_{3/2}$-$1p_{1/2}$} states for a generic $A=44$ system using a finite nuclear potential and observed a substantial decrease in the splitting as the $1p_{1/2}$ moved close to threshold. This demonstrated that while such effects are most pronounced for zero angular momentum neutrons, they are also significant for an angular momentum of one unit for which the centrifugal barrier is still relatively small, at $\sim$500~keV.

%---------------------------------------------------FIGURE 2------------------------------------------------------
\begin{figure}
\centering
\includegraphics[scale=0.8]{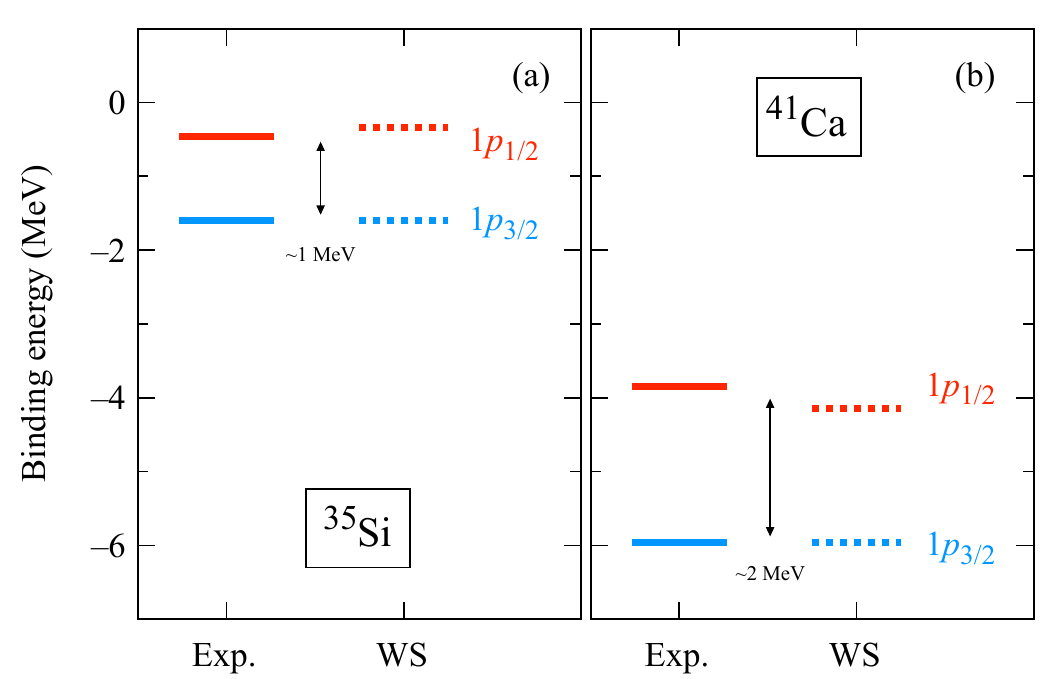}
\caption{\label{fig2} For (a) $^{35}$Si and (b) $^{41}$Ca, a comparison of the experimentally determined binding energies (Exp.)~\cite{note} of the  $1p$ orbitals with those obtained from Woods-Saxon calculations (WS) with a fixed spin-orbit potential, potential depths of 47.0~MeV ($^{35}$Si) and 51.8~MeV ($^{41}$Ca), and parameters given in the text. }
\end{figure}
%-----------------------------------------------------------------------------------------------------------------------

In the following, we will discuss how the changes observed experimentally for $1p_{3/2}$ and $1p_{1/2}$ states from Ca to Si can be described in terms of the proximity of the $p_{1/2}$ orbital to the neutron threshold. To explore such effects, calculations were carried out with a Woods-Saxon potential using the code of Volya~\cite{Volya} and several established parameter sets characterizing the neutron-nucleus potential. In Fig.~\ref{fig2} we show the binding energies of the $1p$ levels from experiment for $^{34}$Si$+n$ and $^{40}$Ca$+n$ and from Woods-Saxon calculations with potential parameters  $r_0=1.28$~fm, $a=0.63$~fm, $r_{\rm so0}=1.1$, $a_{\rm so}=0.65$~fm, and $V_{\rm so}=6$~MeV (as used in Ref.~\cite{Hoffman14}). The depth of the potential was chosen to reproduce the binding of the $1p_{3/2}$ orbital. Note, the spin-orbit potential is the same for both Si and Ca.

Immediately it can be seen that the general feature, a decrease of $\sim$1~MeV in the separation of the $1p_{3/2}$ and $1p_{1/2}$ states, is reproduced by the calculations without any change to the spin-orbit strength. At $^{35}$Si the $1p_{1/2}$ orbital is just bound by a few hundred keV. With no experimental information yet available on the fragmentation of the $1p$ states, it is possible the $1p$ are slightly less bound as fragmentation would, most likely, shift the centroid that way.

Other Woods-Saxon calculations using a range of sensible parameters, for example, $r_0=1.25$--1.28~fm, $a=0.60$--0.75~fm, $r_{\rm so0}=1.1$, $a_{\rm so}=0.65$--0.80~fm, and $V_{\rm so}=6$--7.5~MeV, were explored. In these cases, the depth of the binding potential chosen to reproduce the experimental binding energy of the $1p_{3/2}$ orbital. Typical potential depths for $^{40}$Ca$+n$ were around 52 MeV and for $^{34}$Si$+n$, 47 MeV, consistent with the global parameterization of Ref.~\cite{Schwierz07}. For some of the parameter sets, the $1p_{1/2}$ is just slightly bound (by a few hundred keV) and for others slightly unbound. The {\it change} in the separation between the $1p_{3/2}$ and $1p_{1/2}$ spin-orbit partners from $^{41}$Ca to $^{35}$Si is relatively insensitive to the choice of the parameter, varying from around 0.7--0.9~MeV. The relative proximity of these orbitals to the separation threshold is what dominates the change in separation between the $1p$ orbitals, without invoking a weakening of the spin-orbit force.

%---------------------------------------------------FIGURE 3------------------------------------------------------
\begin{figure}
\centering
\includegraphics[scale=0.8]{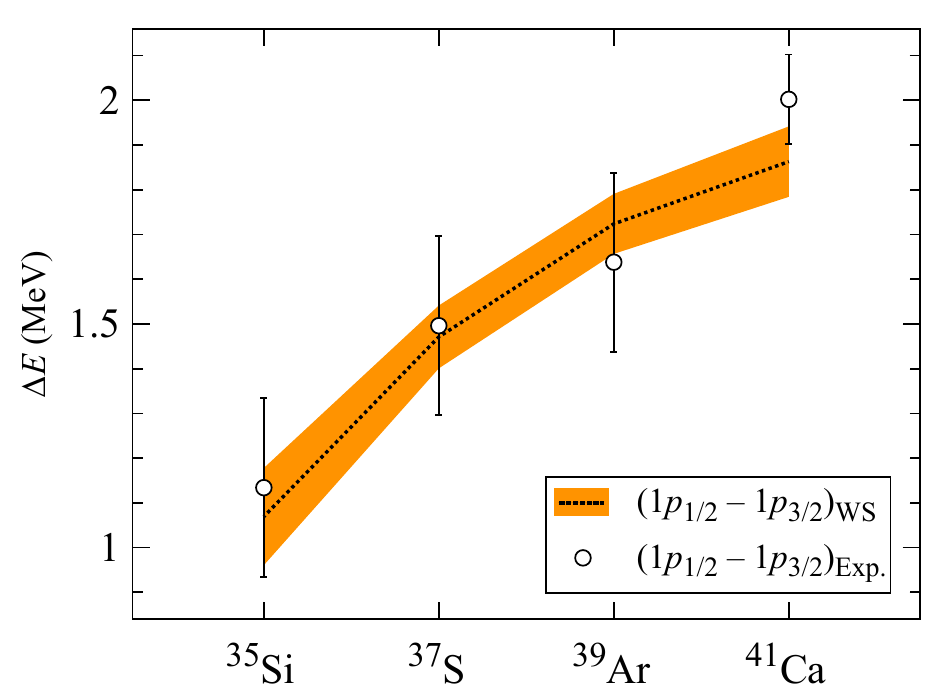}
\caption{\label{fig3} A comparison between experimental spin-orbit splitting of the $1p$ states at $N=21$ for $14\leq Z\leq20$ compared with calculations of the same splittings in a Woods-Saxon potential with a fixed spin-orbit strength. The width of the shaded region is to give a measure of the uncertainties associated with the calculations. The uncertainties on the experimental data points are discussed in Ref.~\cite{note}.}
\end{figure}
%-----------------------------------------------------------------------------------------------------------------------

Figure~\ref{fig3} shows the difference in binding energies of the $1p_{3/2}$ and $1p_{1/2}$ neutron single-particle orbitals from the experimental data and from the Woods-Saxon calculations described above. The shaded band showing the theoretical calculations is an indication of the range of uncertainties associated with the choice of reasonable Woods-Saxon parameters. The parameters characterizing the spin-orbit interaction were fixed, being the same for all nuclei in the calculations.

Key to all calculations is that the $1p_{1/2}$ orbital approaches the neutron separation threshold in a manner similar to the experimental data. As the state approaches threshold it moves more slowly with changing potential well depth and/or changing potential radius. The $1p_{3/2}$ orbital is more deeply bound and it moves more rapidly in energy as the nucleus gets smaller. Since the dominant features of the experimental data are quantitatively described by the consequences of finite-binding effects on the $1p_{1/2}$ orbital, this effect must be taken into account before discussing any changes in the spin-orbit interaction strength.

The lingering below threshold is associated with extended rms radii. While bound states with higher-$\ell$ values have rms radii more in line with a typical $r_0A^{1/3}$ dependence, as low-$\ell$ orbits approach threshold their wave functions also become significantly extended---the centrifugal barrier for $\ell=1$ neutrons is only a few hundred keV in these cases.

From the same calculations, the separation between the $0f_{7/2}$ and $0f_{5/2}$ states, while much larger~\cite{note2}, also decreases by about 1~MeV from $^{41}$Ca to $^{35}$Si. This is a consequence of the same mechanism. Although these higher-$\ell$ orbitals experience a larger centrifugal barrier, they do behave similarly at their respective barriers. The $0f_{5/2}$ is unbound by at least 3~MeV at $^{35}$Si, which is within an MeV or so of the barrier. Comparisons between calculations and experimental data for the $0f$ states are complicated by the fact that the $0f_{5/2}$ states are fragmented and in most cases unbound.

Two-body matrix element calculations carried out to determine the residual interaction strength between $\pi 0d_{3/2}$--$\nu1p_{1/2,3/2}$ and $\pi 1s_{1/2}$--$\nu1p_{1/2,3/2}$ orbitals (using the effective interaction from Ref.~\cite{st}) suggest that the splitting between the $1p_{3/2}$ and $1p_{1/2}$ due to the proton-neutron interaction changes by about a hundred keV from Si to Ca, about 10-20\% of the weak-binding effect demonstrated in this work. 

The decrease in the separation between $1p_{3/2}$ and $1p_{1/2}$ orbitals was considered as indirect evidence for a so-called ``bubble'' nucleus at $^{34}$Si~\cite{Mutschler17}. Without direct evidence for a proton bubble at $^{34}$Si, two arguments were put forward. Proton-removal reactions demonstrated that at $^{34}$Si the proton $1s_{1/2}$ orbital is approximately empty. Similarly, from proton-removal transfer reactions on $^{36}$S, the same orbital is essentially fully occupied. While neither reaction probes the charge density directly, the inferred lower density at the center of $^{34}$Si is assumed to contribute significantly to the overall magnitude of the spin-orbit interaction. The threshold effects reported in the present work suggest that such an explanation for the reduction of the spin-orbit strength might not be needed, and that the proximity of the threshold can explain the reduction of the splitting between these spin-orbit partners.

Recent calculations~\cite{Karakatsanis17} using covariant density functionals seem to reproduce the magnitude in the separation of the $1p$ states, however, this seems dominantly to be the result of a sudden decrease in the binding of the $1p_{3/2}$ orbital while the binding of the $1p_{1/2}$ orbital decreases almost linearly with $Z$ (see Fig.~8 of Ref.~\cite{Karakatsanis17}). This is in contrast to what would be expected as the levels approach threshold in a finite nuclear potential. Their single-particle solutions are expanded in an oscillator basis, and would not be sensitive to the geometrical effects discussed above. The change they observe is attributed to a weakening of the spin-orbit interaction. The {\it ab inito} calculations of Ref.~\cite{Duguet17} again suggests a weakening of the $1p$ splitting, but the calculations expand the one-, two-, and three-body operators in a harmonic oscillator basis.

%%---------------------------------------------------FIGURE 4------------------------------------------------------
%\begin{figure}
%\centering
%\includegraphics[scale=0.8]{fig4.pdf}
%\caption{\label{fig4} Comparing the experimentally determined difference (Exp.) between the $1p_{3/2}$ and $1p_{1/2}$ orbitals at $^{41}$Ca  and $^{35}$Si with calculations showing one-body mean-field (WS) and two-body~\cite{st} contributions. For the effective two-body contributions, the total is broken down to show central, tensor, and spin-orbit components. }
%\end{figure}
%%-----------------------------------------------------------------------------------------------------------------------

It is interesting to note that a similar situation might be encountered at $N=17$ going from $^{32}$S to $^{30}$Si. While proton removal reactions~\cite{nndc} from these two nuclei clearly indicate a reduction of the proton $1s_{1/2}$ occupancy in $^{30}$S, charge density distributions, derived from elastic electron scattering data~\cite{deVries}, do not seem to support the appearance of a bubble. However, such data are perhaps not sufficiently precise to discern a central depletion in charge density.

We can also look at the same spin-orbit splittings at \mbox{$N=29$}, where the $1p_{3/2}$ and $1p_{1/2}$ orbitals are relatively well bound at $^{49}$Ca. Here the experimental data are more limited as there are no stable $N=28$ isotones below $^{48}$Ca. However, the same Woods-Saxon calculations described above support the experimental observation of a $p$-wave, almost certainly a $p_{3/2}$, halo at $^{37}$Mg at $N=25$~\cite{Kobayahi14}, a consequence of the same weak-binding mechanism.

In conclusion, we show that simple geometric effects describe the $1p_{3/2}$-$1p_{1/2}$ splitting outside of $N=20$ {\it without} invoking a weakening of the one-body spin-orbit interaction or an additional explicit two-body spin-orbit interaction. 

Previous calculations (e.g., Refs.~\cite{Karakatsanis17,Duguet17}) exploring these trends at $N=21$ used oscillator wave functions, and could not capture the effects of finite binding. A simple Woods-Saxon potential appears to describe a huge variety of near threshold, non-linear, phenomena such as $s$- and $p$-wave halos, the ordering of $sd$ states in light nuclei, and the possible onset of halos or neutron skins around $^{78}$Ni or below $^{208}$Pb at $N=126$~\cite{Hoffman14}. It would be interesting to have better experimental data on the centroids of the neutron $d$ states for the $N=17$ and 19 nuclei, where no change would be expected in the separation of the centroids, and also ($e$,$e$) scattering data on these radioactive nuclei. $^{34}$Si happens to represent the perfect case where the binding energy of the $p_{1/2}$ is very small due to geometrical effects of the finite nuclear potential, something that is not captured in most shell-model calculations. 

We thank John Schiffer and Olivier Sorlin for their insightful comments. AOM also acknowledges discussions with the Berkeley nuclear structure group. This material is based upon work supported by the U.S. Department of Energy, Office of Science, Office of Nuclear Physics, under Contract Numbers DE-AC02-06CH11357 (ANL) and DE-AC52-07NA27344 (LBNL).

%-----------------------------------------------------------------------------------------------------------------------


\begin{references}
\bibitem{note} The centroids of the single-particle strength are the spectroscopic-factor weighted energies of the fragments. For all, except $^{35}$Si where only the single states are known~\cite{Burgunder14}, we only used states with firm spin-parity assignments in the Evaluated Nuclear Structure Data File Database~\cite{nndc}, and  references therein. Estimated uncertainties are of the order of a few hundred keV  for $^{39}$Ar and $^{37}$S and around 50~keV for $^{41}$Ca. Those for $^{35}$Si are difficult to estimate, because they depend on possible missed states, but we adopt an uncertainty of $\pm$200~keV on the $1p_{1/2}$-$1p_{3/2}$ splitting.
\bibitem{note2} The observed value of $\approx$6~MeV is consistent with the expected scaling of a factor of 3 compared to the $1p_{3/2}$-$1p_{1/2}$ splitting in $^{41}$Ca.
\end{references}
\end{document}